\documentclass{iopjournal}
\usepackage{amsmath}
\usepackage{amssymb}
\usepackage{ragged2e}
\usepackage{indentfirst}
\usepackage{cite}
\begin{document}

\articletype{Paper} 

\title{Theoretical Study for Generating Optical GKP State via a Single-Photon-Added Squeezed Vacuum
}

\author{Deriyan Senjaya$^{1,*}$\orcid{0000-0000-0000-0000}}

\affil{$^1$Department of Physics, National Tsing Hua University, Hsinchu 30013, Taiwan}

\affil{$^*$Corresponding Author}

\email{\url{d_senjaya@gapp.nthu.edu.tw}}

\begin{abstract}
\justifying \noindent A theoretical framework is developed to analyze the generation of the optical GKP state using a single‑photon–added squeezed vacuum. This state, defined by the squeezing parameter $r$, is injected into a 50:50 beam splitter, and the optical GKP state is obtained through conditional measurement at one output port. The single‑photon–added squeezed vacuum is especially prominent in this context because it provides a simpler and more experimentally accessible ingredient than Schrödinger cat states, while conditional measurement ensures projection onto a state that closely approximates the finite‑energy GKP form. Fidelity is employed to quantify this closeness, and the analysis demonstrates that the scheme achieves a maximum fidelity of 85\%  at a squeezing level of $3.76 \ \text{dB}$. This performance surpasses approaches based on squeezed optical odd Schrödinger cat states, underscoring the single‑photon–added squeezed vacuum as a practical and effective pathway toward fault‑tolerant photonic quantum computing.
\end{abstract}


\section{Introduction}
 \justifying Quantum mechanics has become a cornerstone of modern technological innovation, with photonic quantum computing emerging as a particularly promising direction. Unlike classical computers, which encode information in binary form (0s and 1s) \cite{Brookshear2007}, photonic quantum computers encode information in the quantum states of light—specifically, single photons—known as photonic qubits \cite{Adami1998}. The principle of superposition allows these qubits to exist in linear combinations of basis states, $|\psi\rangle=c_{\text{0}}|0\rangle+c_{\text{1}}|1\rangle$, thereby enabling computational capabilities beyond classical limits \cite{Suominen2005}. Their evolution is governed by optical quantum gates, represented as unitary operators $\hat{U}$ satisfying $\hat{U}^{\dagger}=\hat{U}^{-1}$ \cite{Suominen2005, Nielsen2010}, which preserve quantum probability density in accordance with Born’s rule.
 

In practice, however, quantum systems are highly susceptible to decoherence. Environmental interactions and imperfections in optical gates disrupt superposition, leading to information loss \cite{Nielsen2010}. Effective error correction is therefore essential. Classical error correction relies on bit repetition, but quantum systems cannot employ this strategy due to the No-Cloning Theorem \cite{Mermin2001}. As a result, photonic quantum computers require more sophisticated schemes. Among these, the Gottesman–Kitaev–Preskill (GKP) state has proven especially promising \cite{Gottesmann2001, Samuel2005, Weedbrook2012}, as it encodes qubits into bosonic fields and enables efficient error correction.


 Despite its theoretical appeal, realizing the optical GKP state remains experimentally challenging. Preparing suitable quantum state ingredients and mitigating environmental noise are significant obstacles \cite{Shi2019, Hastrup2022, Dahan2023}. Vasconcelos \cite{Vasconcelos2010} proposed constructing the optical GKP state using Schrödinger cat states combined with a 50:50 beam splitter and conditional measurement. While theoretically sound, generating cat states experimentally is demanding. To address this, subsequent work \cite{Parigi2007, Bellini2010, Yiru2024} introduced single-photon addition to squeezed vacuum states, achieving high fidelity relative to the odd Schrödinger cat state.


 These developments highlight both the promise and the difficulty of realizing optical GKP states. The relative simplicity of the beam-splitter approach and the improved feasibility of single-photon-added squeezed vacua motivate further theoretical investigation. The present study builds on these methods to explore the feasibility of constructing robust optical GKP states, with the aim of identifying key parameters that may accelerate progress toward fault-tolerant photonic quantum computing.

\section{Theoretical Basis of GKP State and the Optical GKP State}

\begin{figure}[h]
    \centering
    \includegraphics[width=1\linewidth]{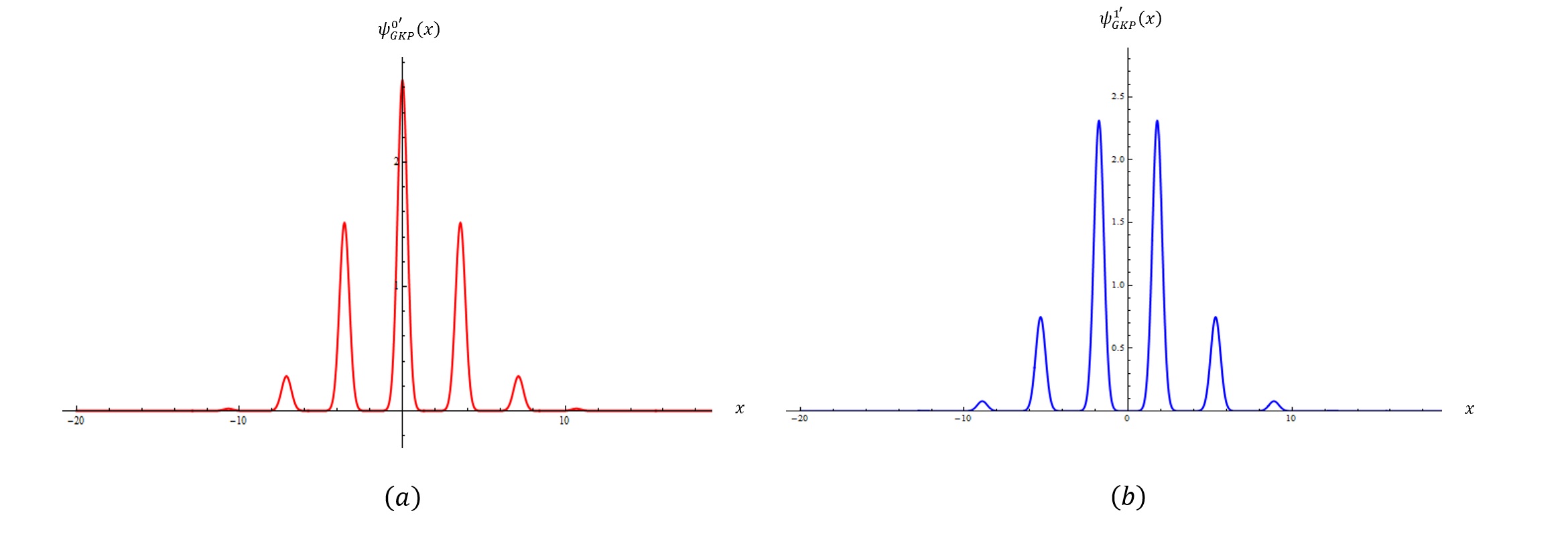}
    \caption{Physical GKP states from the Gaussian envelope approximation: (a) $\psi_{\text{GKP}}'^{0}(x)$ and (b) $\psi_{\text{GKP}}'^{1}(x)$}
    \label{fig:1}
\end{figure}

The Gottesman–Kitaev–Preskill (GKP) state is generally regarded as a hybrid quantum state that integrates discrete variables (qubits) with continuous variables, encoded in bosonic fields \cite{Gottesmann2001, Grismo2021}. In this framework, qubits are embedded into continuous-variable systems, making the GKP state a member of the broader class of continuous-variable quantum states. Mathematically, the GKP state $\psi_{\text{GKP}}(x)$ is defined as the eigenfunction of a specific eigenvalue relation \cite{Conrad2021, Rymarz2021}, with the convention $\hbar=1$.


\begin{equation}
\label{1}
    \hat{H}_{\text{GKP}}\psi_{\text{GKP}}(x)=E_{\text{GKP}}\psi_{\text{GKP}}(x)
\end{equation}
\begin{equation}
\label{2}
    \left[-\frac{1}{2}J_{0}(\hat{S}_{X}^{\dagger}+\hat{S}_{X}+\hat{S}_{Z}^{\dagger}+\hat{S}_{Z})\right]\psi_{\text{GKP}}(x)=E_{\text{GKP}}\psi_{\text{GKP}}(x)
\end{equation}
Here, $J_{0}$ denotes a coupling constant with units of energy, while $\hat{S}_{i}$ ($i = X, Z$) represents the stabilizer operators. The GKP state $\psi_{\text{GKP}}(x)$ satisfies Eq.(\ref{2}) because it is simultaneously an eigenfunction of both stabilizers $\hat{S}_{X}$ and $\hat{S}_{Z}$. Moreover, the formulation in Eq.(\ref{2}) exhibits two-fold degeneracy, indicating that two distinct GKP states correspond to the same energy eigenvalue. These degenerate states are explicitly expressed in Eq.(\ref{3}).

\begin{equation}
\label{3}
    \psi_{\text{GKP}}(x) = \begin{cases}
  \psi_{\text{GKP}}^{0}(x)=\sum^{\infty}_{j=-\infty}\delta(x-2j\sqrt{\pi}) &  \forall j\in \mathbb{Z} \\
  \psi_{\text{GKP}}^{1}(x)=\sum^{\infty}_{j=-\infty}\delta(x-(2j+1)\sqrt{\pi}) & \forall j\in \mathbb{Z}
\end{cases}
\end{equation}
In this context, $\mathbb{Z}$ denotes the set of integers $(\dots,-1,0,1,\dots)$. 
Eq.(\ref{3}) represents a Dirac comb with a period of $2\sqrt{\pi}$. 
However, GKP states expressed in the form of a Dirac comb are nonphysical because they cannot be normalized \cite{Grismo2021}. 
To render these states physically meaningful (i.e., normalizable), an approximation method is introduced. 
This approach applies a Gaussian envelope, which ensures that the GKP states become properly normalized \cite{Grismo2021}.

\begin{equation}
    \label{4}
    \psi'_{\text{GKP}}(x)=\iint dUdV \ \eta_{\Delta}(U,V) e^{\frac{1}{2}iUV}e^{-iU\hat{p}}e^{iV\hat{x}}\psi_{\text{GKP}}(x)
\end{equation}
In Eq.(\ref{4}), $\Delta$ denotes the width of the Gaussian envelope, while $\eta_{\Delta}(U,V) \approx \frac{1}{\pi \Delta^{2}} 
\exp\!\left(-\frac{1}{2\Delta^{2}}(U^{2}+V^{2})\right)$ 
represents the error function introduced by the envelope, valid for $0 < \Delta < 1$ \cite{Grismo2021}. 
By substituting Eq.(\ref{3}) into Eq.(\ref{4}) and carrying out the integration, 
the physical GKP states are obtained, as presented in Eq.(\ref{5}) and Eq.(\ref{6}), 
and illustrated in Fig.\ref{fig:1}.

\begin{equation}
    \label{5}
    \psi_{\text{GKP}}'^{0}=\sqrt{\frac{2}{\pi\Delta^{2}}}\sum^{\infty}_{j=-\infty}e^{-\frac{1}{2}\Delta^{2}(4j^{2}\pi)}\left[e^{-\frac{1}{2\Delta^{2}}(x-2j\sqrt{\pi})^{2}}e^{-\frac{1}{2\Delta^{2}}\left(\frac{1}{4}(x-2j\sqrt{\pi})^{2}-2(x-2j\sqrt{\pi})j\sqrt{\pi}\right)}\right] \ \forall j \in \mathbb{Z}
\end{equation}

\begin{equation}
    \label{6}
    \begin{split}
         \psi_{\text{GKP}}'^{1}=\sqrt{\frac{2}{\pi\Delta^{2}}}\sum^{\infty}_{j=-\infty}e^{-\frac{1}{2}\Delta^{2}((2j+1)^{2}\pi)} e^{-\frac{1}{2\Delta^{2}}(x-(2j+1)\sqrt{\pi})^{2}} \\ 
         e^{-\frac{1}{2\Delta^{2}}\left(\frac{1}{4}(x-(2j+1)\sqrt{\pi})^{2}-(x-(2j+1)\sqrt{\pi})(2j+1)\sqrt{\pi}\right)} \ \forall j \in \mathbb{Z}
    \end{split}
\end{equation}

The states $\psi_{\text{GKP}}'^{0}(x)$ and $\psi_{\text{GKP}}'^{1}(x)$ can serve as basis states for continuous-variable quantum systems. 
These basis states can be transformed into alternative forms, such as $\psi_{\text{GKP}}^{+}(x)$ and $\psi_{\text{GKP}}^{-}(x)$. Although expressed differently, these new bases remain variations of the GKP states, since they continue to satisfy the eigenvalue relation in Eq.(\ref{2}). The constants $N_{\text{GKP}}^{\pm}$ provide the necessary normalization for these states.

\begin{equation}
    \label{7}
    \psi_{\text{GKP}}^{\pm}(x)=N_{\text{GKP}}^{\pm}\left(\psi_{\text{GKP}}^{'0}(x)\pm\psi_{\text{GKP}}^{'1}(x)\right)
\end{equation}
The optical GKP state also satisfies the mathematical definition given in Eq.(\ref{2}). The term \emph{optical} arises from the specific construction of the stabilizer operators $\hat{S}_{i}$ ($i = X, Z$). 
Within the optical GKP framework, these stabilizer operators are defined using two distinct displacement operators, $\hat{D}(\alpha)$ and $\hat{D}(\beta)$ \cite{Grismo2021}. These displacement operators obey the Weyl--Heisenberg relation shown in Eq.(\ref{8}), subject to the condition $\beta \alpha^{*} - \beta^{*} \alpha = i\pi$ \cite{Grismo2021, Serafini2017}.

\begin{equation}
    \label{8}
    \hat{D}(\alpha)\hat{D}(\beta)-\hat{D}(\beta)\hat{D}(\alpha)e^{-(\beta\alpha^{*}-\beta^{*}\alpha)}=0
\end{equation}
The condition $\beta \alpha^{*} - \beta^{*} \alpha = i\pi$ allows Eq.(\ref{8}) to be expressed as the anti-commutation relation 
$\{\hat{D}(\alpha), \hat{D}(\beta)\} = 0$. This anti-commutation mirrors the relation between the Pauli operators $\hat{\sigma}_{X}$ and $\hat{\sigma}_{Z}$, 
where $\{\hat{\sigma}_{X}, \hat{\sigma}_{Z}\} = 0$. Consequently, one can identify $\hat{\sigma}_{X} = \hat{D}(\alpha)$ and $\hat{\sigma}_{Z} = \hat{D}(\beta)$. With this correspondence, the stabilizers of the optical GKP state can be formally defined as shown in Eq.(\ref{9}) and Eq.(\ref{10}).

\begin{equation}
    \label{9}
    \hat{S}_{X}=\hat{\sigma}_{X}\hat{\sigma}_{X}=\hat{D}(2\alpha)
\end{equation}
\begin{equation}
    \label{10}
    \hat{S}_{Z}=\hat{\sigma}_{Z}\hat{\sigma}_{Z}=\hat{D}(2\beta)
\end{equation}

\section{Candidate of The Optical GKP State via Single-Photon Added Squeezed Vacuum}
\label{candidate}
Sec.\ref{candidate} focuses on the theoretical derivation of a candidate optical GKP state generated through a single-photon–added squeezed vacuum. The starting point is the definition of the squeezed vacuum. A squeezed vacuum is obtained by applying the squeezed operator $\hat{S}(\xi)$ to the ground state of the quantum harmonic oscillator, $\psi_{0}(x)$, yielding $\psi_{\xi}(x) = \hat{S}(\xi)\psi_{0}(x)$.


\begin{equation}
    \label{11}
    \psi_{\xi}(x)=\frac{1}{\sqrt{\cosh r}}\sum^{\infty}_{n=0}\left[\frac{(-1)^{2n}\sqrt{(2n)!}}{2^{n}n!}e^{in\phi}\tanh^{n}r\right]\psi_{2n}(x)
\end{equation}
Here, $\psi_{2n}(x) = \frac{1}{\sqrt{2^{n} n!}} H_{2n}(x) e^{-\tfrac{1}{2}x^{2}}$ denotes the eigenfunction of the quantum harmonic oscillator in the $2n$-state, where $H_{2n}(x)$ is the Hermite polynomial of order $2n$. 
By applying the creation operator $\hat{a}^{\dagger}$ to Eq.(\ref{11}), namely $\hat{a}^{\dagger}\psi_{\xi}(x)$, the single-photon–added squeezed vacuum is obtained, as expressed in Eq.(\ref{12}).

\begin{equation}
    \label{12}
    \psi_{\bar{\xi}}(x)=\frac{1}{\sqrt{I_{0}}}\sum^{\infty}_{n=0}\left[\frac{(-1)^{n}}{2^{n}n!}\sqrt{(2n+1)!}e^{in\phi}\tanh^{n}r\right]\psi_{2n+1}(x)
\end{equation}
The factor $\frac{1}{\sqrt{I_{0}}}$ serves as the normalization constant in Eq.(\ref{12}), $I_{0} = \sum_{n=0}^{\infty} \frac{1}{2^{2n}(n!)^{2}} \tanh^{2n} r$ defines the normalization term. Meanwhile, $\psi_{2n+1}(x)$ denotes the eigenfunction of the quantum harmonic oscillator in the $(2n+1)$-state.

\begin{figure}[h]
    \centering
    \includegraphics[width=1\linewidth]{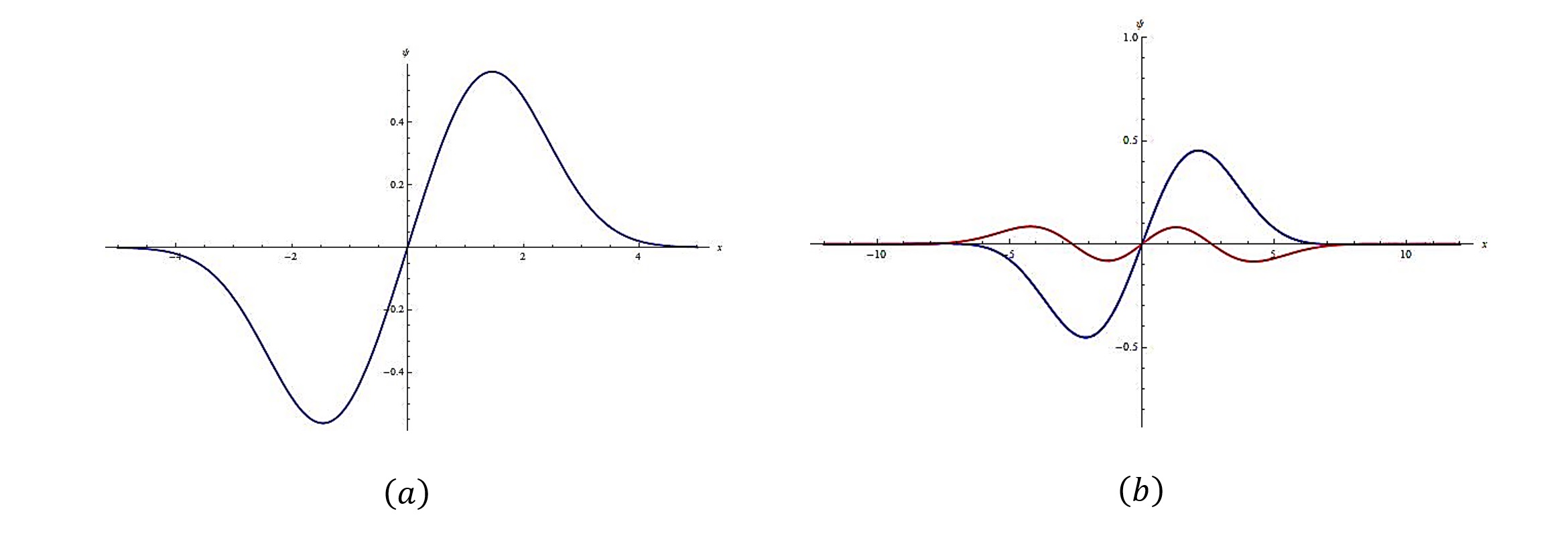}
    \caption{Comparison between (a) the optical odd-Schrodinger cat state and (b) single-photon added squeezed vacuum with $\phi\neq0 \vee \pi$ (In this case $\phi=\pi/3$, blue is the real part, and red is the imaginary part)}
    \label{fig:2}
\end{figure}

The single-photon–added squeezed vacuum described in Eq.(\ref{12}) can serve as an alternative to the optical Schrödinger cat state. 
However, it can only approximate the optical odd Schrödinger cat state for specific values of the parameters $r$ and $\phi$ \cite{Vasconcelos2010,Yiru2024}. 
The optical odd Schrödinger cat state is defined as a linear combination of two coherent states, 
$\psi_{\alpha}(x) = \psi_{0}(x - \alpha\sqrt{2})$ and $\psi_{-\alpha}(x) = \psi_{0}(x + \alpha\sqrt{2})$, 
combined with a negative  sign $\psi^{-}_{\text{SC}}(x) = \left(2\left(1 - e^{-2\alpha^{2}}\right)\right)^{-1/2} 
\left(\psi_{\alpha}(x) - \psi_{-\alpha}(x)\right)$ \cite{Ourjoumtsev2008,Takase2021}. If the parameter $\phi$ differs from $0$ or $\pi$, the single-photon–added squeezed vacuum acquires an imaginary component, making it distinct from the optical odd Schrödinger cat state, as illustrated in Fig.\ref{fig:2}.

\begin{figure}[ht]
    \centering
    \includegraphics[width=1\linewidth]{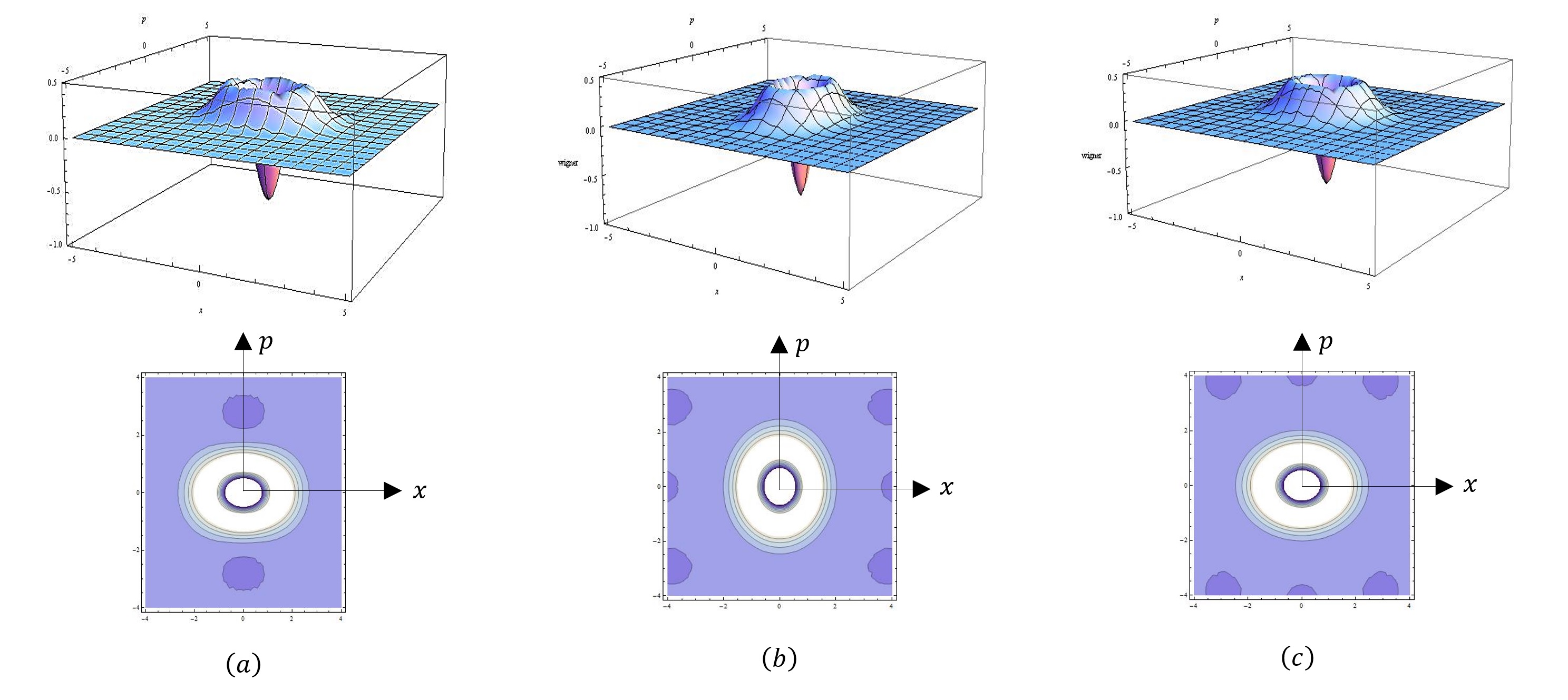}
    \caption{The Wigner function for (a) the optical odd-Schrodinger cat state ($\alpha=0.8$), (b) the single-photon added squeezed vacuum with $\phi=0$ ($r=0.1$), and (c) the single-photon added squeezed vacuum with $\phi=\pi$ ($r=0.1$)}
    \label{fig:3}
\end{figure}

The cases $\phi = 0$ and $\phi = \pi$ yield similar outcomes, which introduces a potential ambiguity. 
To resolve this issue, one can compare their corresponding Wigner functions. The Wigner functions of the optical odd Schrödinger cat state and the single-photon–added squeezed vacuum are given in Eq.(\ref{13}) and Eq.(\ref{14}).

\begin{equation}
    \label{13}
    \begin{split}
          W_{\text{SC}}^{-}(p,x) &=\frac{2}{\pi}\sqrt{\frac{\pi}{2}}\left[e^{-(x-\alpha\sqrt{2})^{2}-p^{2}}+e^{-(x+\alpha\sqrt{2})^{2}-p^{2}}\right] \\ &+ \frac{2}{\pi}\sqrt{\frac{\pi}{2}}\left[-e^{-(x^{2}+2\alpha^{2})}e^{-(p+i\alpha\sqrt{2})^{2}}+e^{-(x^{2}+2\alpha^{2})}e^{-(p-i\alpha\sqrt{2})^{2}}\right]
    \end{split}
\end{equation}
\begin{equation}
    \label{14}
    W_{\bar{\xi}}(p,x)=\frac{1}{2\pi I_{0}\sqrt{2}}\sum^{\infty}_{n,n'=0}\left[\frac{(-1)^{n+n'}}{2^{n+n'}(n!)(n'!)}e^{i(n-n')\phi}\tanh^{n+n'}r\right]d_{n,n'}(p,x)
\end{equation}
In Eq.(\ref{14}), the $d_{n,n'}$ is defined as in Eq.(\ref{15}).

\begin{equation}
    \label{15}
    d_{n,n'}=\int^{\infty}_{-\infty}e^{-ipz}H_{2n+1}\left(x+\frac{1}{2}z\right)H_{2n'+1}\left(x-\frac{1}{2}z\right)e^{-\frac{1}{4}z^{2}}dz
\end{equation}
According to Fig.\ref{fig:3}(b) and Fig.\ref{fig:3}(c), 
the Wigner function in Fig.\ref{fig:3}(c) closely resembles that of Fig.\ref{fig:3}(a), 
primarily due to the matching orientation of the central circle.
The Wigner function analysis in Fig.\ref{fig:3} indicates that the optimal choice of parameter is $\phi = \pi$. Consequently, when applying the method of \cite{Vasconcelos2010} to construct the GKP state, Eq.(\ref{12}) is modified into Eq.(\ref{16}) to serve as the input state.

\begin{equation}
    \label{16}
    \psi_{\bar{\xi}}(x)=\frac{1}{\sqrt{I_{0}}}\sum^{\infty}_{n=0}\left[\frac{(-1)^{n}}{2^{n}n!}\sqrt{(2n+1)!}\tanh^{n}(-r)\right]\psi_{2n+1}(x)
\end{equation}

\begin{figure}[ht]
    \centering
    \includegraphics[width=1\linewidth]{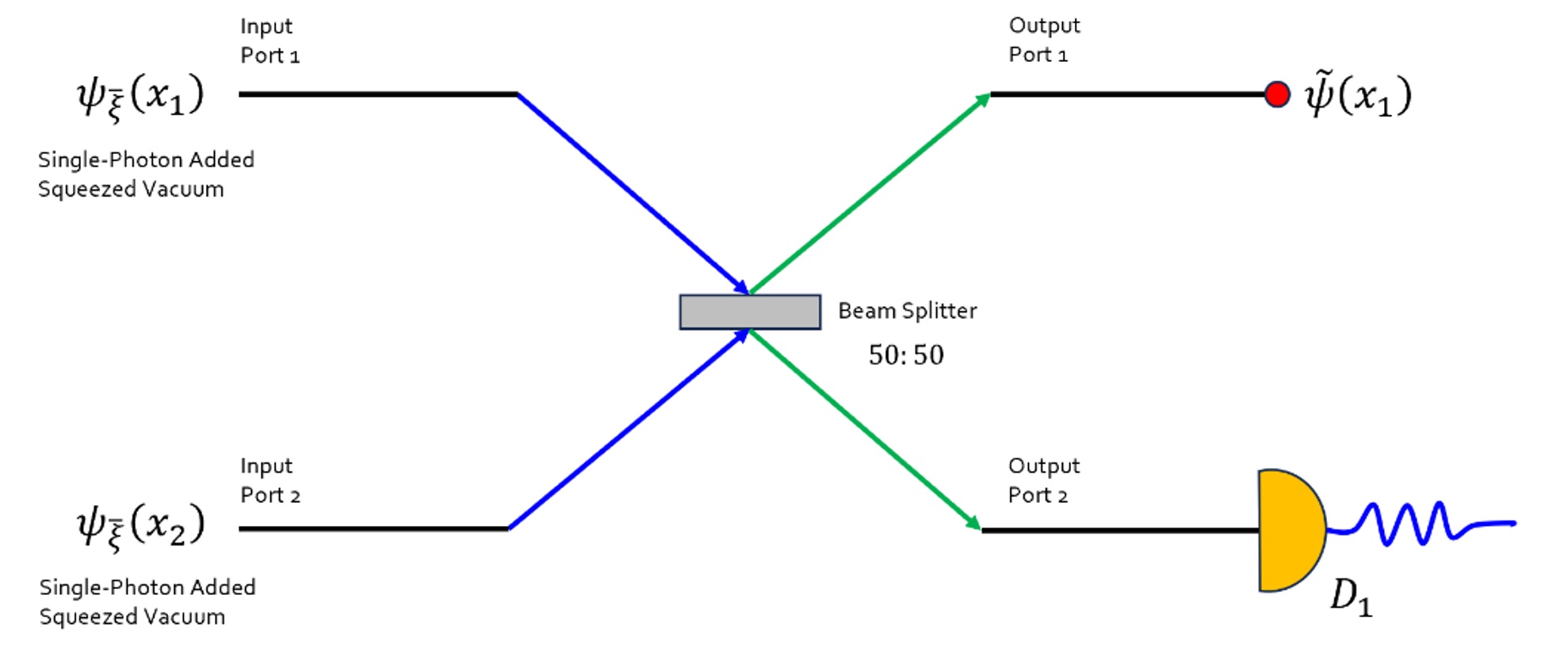}
    \caption{Schematic diagram to create the candidate for the optical GKP state via single-photon added squeezed vacuum.}
    \label{fig:4}
\end{figure}

The method of \cite{Vasconcelos2010} constructs the GKP state by combining two distinct Schrödinger cat states through a 50:50 beam splitter, followed by conditional measurement at one of the output ports. 
In this research, the Schrödinger cat states used in the original Vasconcelos method are replaced with the single-photon–added squeezed vacuum defined in Eq.(\ref{16}). Fig.\ref{fig:4} illustrates the scheme of \cite{Vasconcelos2010} adapted with the single-photon–added squeezed vacuum.

In \cite{Vasconcelos2010}, the first step involves combining the two input states before they enter the beam splitter. 
Mathematically, this combination is expressed as the direct product of the two states $\psi_{\text{in}}(x_{1}, x_{2}) = \psi_{\bar{\xi}}(x_{1}) \times \psi_{\bar{\xi}}(x_{2})$. Eq.(\ref{17}) presents the complete form of this direct product operation between the two states.
\begin{equation}
    \label{17}
    \psi_{\text{in}}(x_{1},x_{2})=\frac{1}{I_{0}}\sum^{\infty}_{n,n'=0}\left[\frac{(-1)^{n+n'}}{2^{n+n'}n!n'!}\sqrt{(2n+1)!(2n'+1)!}\tanh^{n+n'}(-r)\right]\psi_{2n+1}(x_{1})\psi_{2n'+1}(x_{2})
\end{equation}
The second step in \cite{Vasconcelos2010} involves passing the combined input state through a 50:50 beam splitter. In quantum mechanics, the beam splitter is represented by the unitary operator $\hat{U}_{\text{BS}}(\theta)$. For a 50:50 beam splitter, the angle is set to $\theta = \pi/4$. 
The corresponding unitary operator, $\hat{U}_{\text{BS}}(\pi/4)$, is expressed in Eq.(\ref{18}) as a $2 \times 2$ matrix.

\begin{equation}
    \label{18}
    \hat{U}_{\text{BS}}(\pi/4)=\frac{1}{\sqrt{2}}\left[\begin{matrix}
        1 & 1 \\
        1 & -1 \\
    \end{matrix}\right]
\end{equation}
From Eq.(\ref{18}), the beam splitter operator $\hat{U}_{\text{BS}}(\pi/4)$ performs the coordinate transformation $x_{1} \;\rightarrow\; \frac{1}{\sqrt{2}}(x_{1} + x_{2}) = x_{1}';  x_{2} \;\rightarrow\; \frac{1}{\sqrt{2}}(x_{1} - x_{2}) = x_{2}'$. These transformations are equivalent to a rotation of the two-dimensional coordinate frame by an angle $\theta = \pi/4$. A detailed discussion of this coordinate transformation, as the effect of the beam splitter operator $\hat{U}_{\text{BS}}(\pi/4)$, is provided in Appendix A. Accordingly, the output state after the 50:50 beam splitter is given by $\psi_{\text{out}}(x_{1}, x_{2}) = \hat{U}_{\text{BS}}(\pi/4)\,\psi_{\text{in}}(x_{1}, x_{2})$, which can be expressed simply by substituting the old coordinates with the transformed ones, as shown in Eq.(\ref{19}).

\begin{equation}
    \label{19}
    \psi_{\text{out}}(x_{1},x_{2})=\frac{1}{I_{0}}\sum^{\infty}_{n,n'=0}\left[\frac{(-1)^{n+n'}}{2^{n+n'}n!n'!}\sqrt{(2n+1)!(2n'+1)!}\tanh^{n+n'}(-r)\right]\psi_{2n+1}(x_{1}')\psi_{2n'+1}(x_{2}')
\end{equation}

The final step in \cite{Vasconcelos2010} is the conditional measurement. 
When the measurement is performed at the second output port, the process can be expressed mathematically as $\tilde{\psi}(x_{1}) = \int_{-\infty}^{\infty} \psi_{\text{out}}(x_{1}, x_{2}) \, dx_{2}$. By substituting Eq.(\ref{19}) into this integral, one obtains $\tilde{\psi}(x_{1}) = \int_{-\infty}^{\infty} \psi_{\text{out}} \, dx_{2}$, which leads to the result given in Eq.(\ref{20}), with normalization constant $N_{\text{S}}$.

\begin{equation}
    \label{20}
    \tilde{\psi}(x_{1})=\frac{N_{\text{S}}}{\sqrt{\pi}I_{0}}e^{-\frac{1}{2}x_{1}^{2}}\sum^{\infty}_{n,n'=0}\left[\frac{(-1)^{n+n'}}{2^{2n'+2n+1}n!n'!}\tanh^{n+n'}(-r)\right]\varphi_{n,n'}(x_{1})
\end{equation}
In Eq.(\ref{20}), the function $\varphi_{n,n'}(x_{1}) = \int_{-\infty}^{\infty} 
H_{2n+1}(x_{1}') \, H_{2n'+1}(x_{2}') \, e^{-x_{2}^{2}/2} \, dx_{2}$ involves the integral of two Hermite polynomials. However, Eq.(\ref{20}) cannot be directly identified as the optical GKP state, since it contains a single squeezing parameter $r$, which remains adjustable. The parameter $r$ must be tuned such that Eq.(\ref{20}) approaches the form of Eq.(\ref{5}), Eq.(\ref{6}), or Eq.(\ref{7}), 
depending on which expression it most closely resembles. For this reason, Eq.(\ref{20}) is regarded as a \emph{candidate} for the optical GKP state rather than its definitive form.

\section{Result and Discussions}
\label{resdis}
Sec.\ref{resdis} presents two key analyses. 
The first is a preliminary examination of Eq.(\ref{20}), establishing its structural properties and relation to the optical GKP framework. The second analysis determines the optimal squeezing parameter $r$, using fidelity as the criterion for quantifying how closely Eq.(\ref{20}) approximates the GKP state. This fidelity-based evaluation builds upon the insights gained from the preliminary analysis.

\subsection{Preliminary Analysis}
\begin{figure}[ht]
    \centering
    \includegraphics[width=0.9\linewidth]{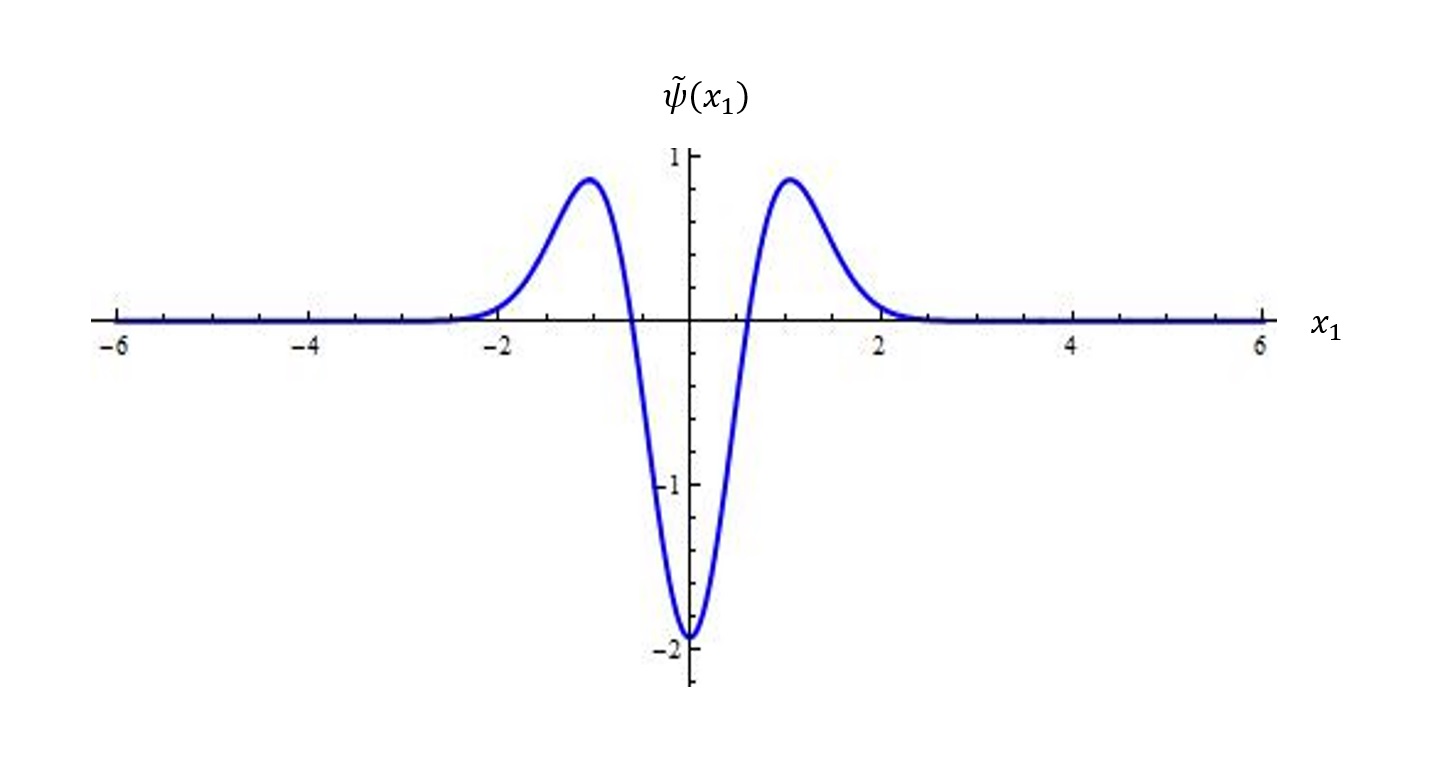}
    \caption{Caption}
    \label{fig:5}
\end{figure}

The preliminary analysis focuses on a rough comparison between Eq.(\ref{20}) and the GKP states expressed in Eq.(\ref{5}), Eq.(\ref{6}), or Eq.(\ref{7}). To carry out this comparison, an arbitrary squeezing parameter is chosen, specifically $r = 0.5$. The graph of Eq.(\ref{20}) is then plotted for indices $n = [0,8]$ and $n' = [0,8]$. The series in Eq.(\ref{20}) is truncated at these values because higher-order terms contribute negligibly to the result. The resulting plot of Eq.(\ref{20}) as a function of the coordinate $x_{1}$ is shown in Fig.\ref{fig:5}.

\begin{figure}[ht]
    \centering
    \includegraphics[width=1\linewidth]{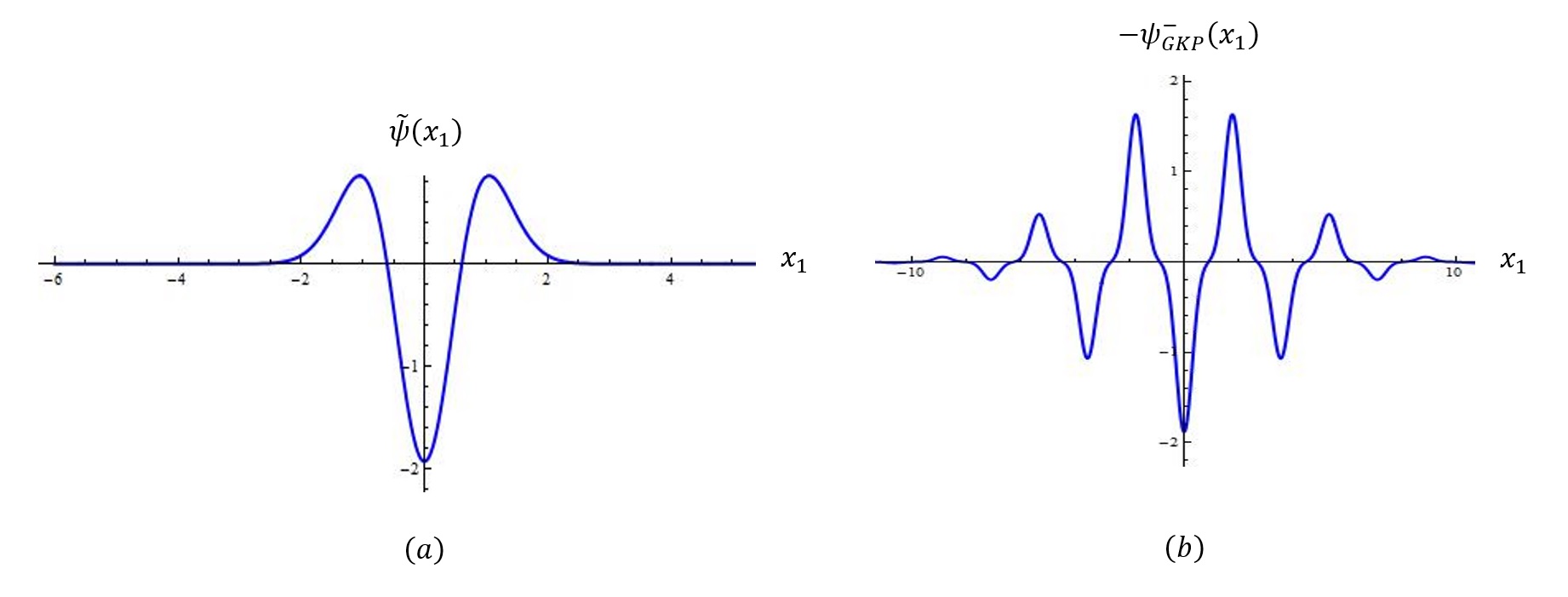}
    \caption{The preliminary morphological comparison between (a) Eq.(\ref{20}) and (b) the negative sign of the GKP state $-\psi_{\text{GKP}}^{-}(x_{1})=-N_{\text{GKP}}^{-}(\psi_{\text{GKP}}^{'0}(x)-psi_{\text{GKP}}^{'0}(x))$ as in Eq.(\ref{7}) with $\Delta=0.1$}
    \label{fig:6}
\end{figure}

\begin{figure}[ht]
    \centering
    \includegraphics[width=1\linewidth]{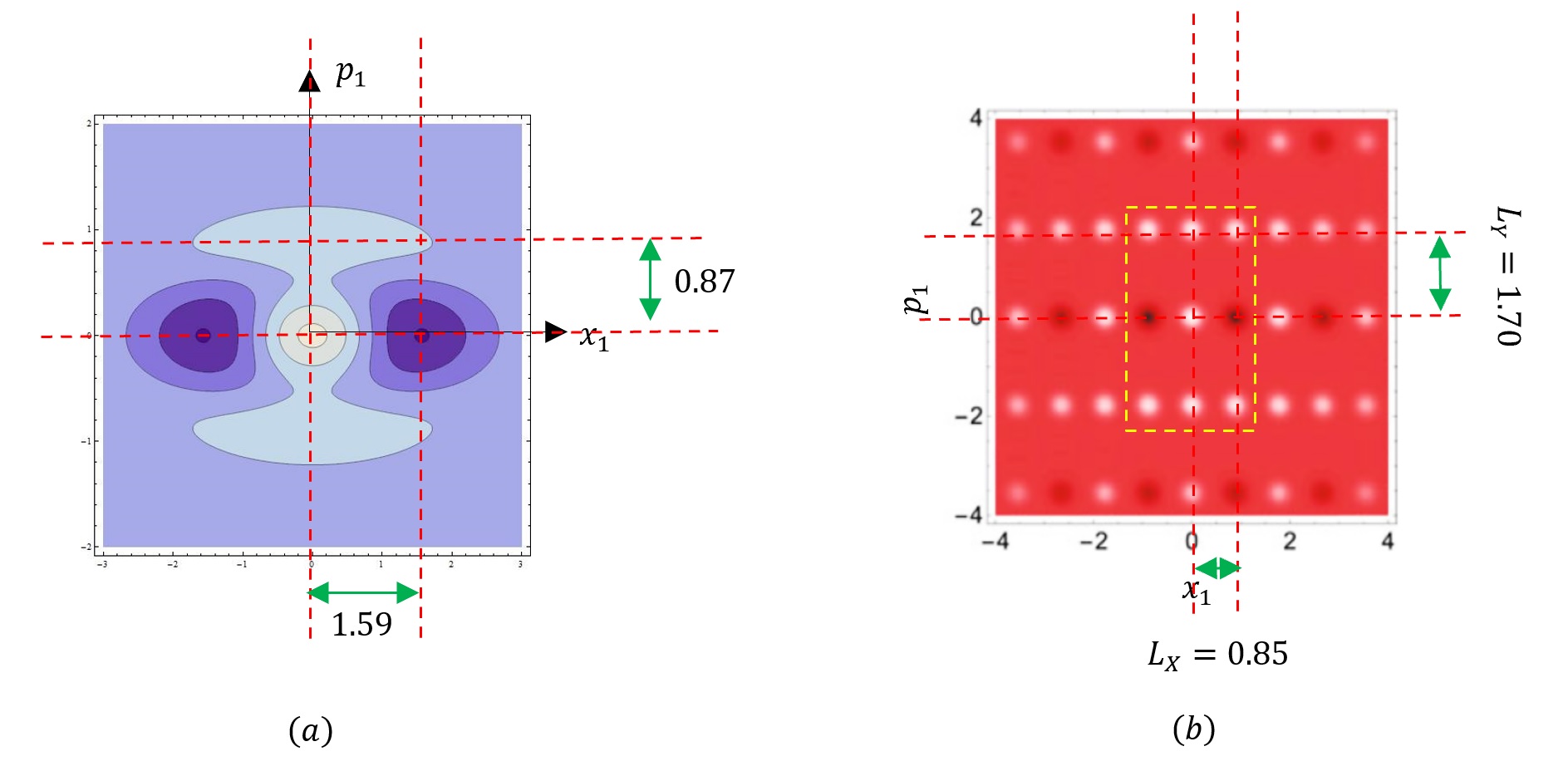}
    \caption{The preliminary morphological comparison of the Wigner function contour plot between (a) Eq.(\ref{20})and (b) $\psi_{\text{GKP}}^{-}$ [16]. In (a), the purple region has a negative value, and the white region has a positive value. The pattern (a) follows (b) inside the yellow box}
    \label{fig:7}
\end{figure}

Fig.\ref{fig:5} demonstrates that Eq.(\ref{20}) most closely resembles the inverse of Eq.(\ref{7}), obtained by introducing a negative sign into Eq.(\ref{7}). However, the number of peaks in Eq.(\ref{7}) must be adjusted by tuning the parameter $\Delta$, as shown in greater detail in Fig.\ref{fig:6}. Furthermore, the identification of Eq.(\ref{20}) as a candidate for the optical GKP state is reinforced by examining its Wigner function, given in Eq.(\ref{21}), 
and the corresponding representation in Fig.\ref{fig:7}.

\begin{equation}
    \label{21}
    \begin{split}
         W_{\tilde{\psi}}(p,x_{1}) &=\frac{N_{\text{S}}^{2}}{\pi I_{0}^{2}\sqrt{2\pi}}\sum^{\infty}_{k,k',n,n'=0}c_{n,n',k,k'} \\ 
         & \cdot \int^{\infty}_{-\infty}e^{-ipz}e^{-\frac{1}{2}(x_{1}+\frac{z}{2})^{2}-\frac{1}{2}(x_{1}-\frac{z}{2})^{2}}\varphi_{n,n'}\left(x_{1}-\frac{z}{2}\right)\varphi_{k,k'}\left(x_{1}+\frac{z}{2}\right)dz
    \end{split}
\end{equation}
In Eq.(\ref{21}), the $c_{n,n',k,k'}$ takes the form:
\begin{equation}
    \label{22}
    c_{n,n',k,k'}=\frac{(-1)^{n+n'+k+k'}}{2^{2n+2n'+2k+2k'}n!n'!k!k'!}\tanh^{n+n'+k+k'}(-r)
\end{equation}
This expression encodes the dependence on the indices $n, n', k, k'$ through factorial terms and alternating signs, 
while the squeezing parameter $r$ enters via the hyperbolic tangent function.

From Fig.\ref{fig:7}(a), the Wigner distribution of Eq.(\ref{20}) exhibits alternating white and purple regions. 
The white regions (corresponding to the white areas in Fig.\ref{fig:7}(b)) represent positive values of the Wigner function, while the purple regions (corresponding to the black areas in Fig.\ref{fig:7}(b)) represent negative values. This alternating pattern is consistent with the reference in \cite{Alvarez2019}, 
although differences arise in the interspacing distance between the regions in Fig.\ref{fig:7}(a) and Fig.\ref{fig:7}(b) along the $x_{1}$ and $p_{1}$ quadratures. The strong similarity in the overall pattern between Fig.\ref{fig:7}(a) and Fig.\ref{fig:7}(b) provides compelling evidence that Eq.(\ref{20}) is indeed a candidate for the optical GKP state. 
The discrepancy in interspacing distance can be corrected by adjusting the squeezing parameter $r$ in Eq.(\ref{20}), with the fidelity parameter $F$ serving as the measure of closeness to the ideal GKP state.

\subsection{Fidelity Analysis}
The fidelity parameter $F$ quantifies the closeness between two distinct quantum states, $\psi_{\text{A}}$ and $\psi_{\text{B}}$ \cite{Suominen2005,Nielsen2010}. 
For the case of two pure states, the fidelity is defined by Eq.(\ref{23}), 
which provides the mathematical expression for evaluating the overlap between $\psi_{\text{A}}$ and $\psi_{\text{B}}$.

\begin{equation}
    \label{23}
    F=\left|\int^{\infty}_{-\infty}\psi_{\text{A}}^{*}(x)\psi_{\text{B}}(x)dx\right|^{2}
\end{equation}
The interpretation of Eq.(\ref{23}) is that it represents the overlap integral between two pure states. In general, the fidelity between two distinct states lies within the range $0 \leq F \leq 1$. When the characteristics of the two pure states are sufficiently similar, their overlap increases, and the fidelity approaches $F = 1$. Conversely, if $F = 0$, the two states are perfectly distinct, i.e., orthogonal. 
Thus, the fidelity parameter serves as a practical guide for determining the optimal squeezing parameter $r$ in Eq.(\ref{20}), 
ensuring that Eq.(\ref{20}) approximates the target GKP state as closely as possible.

\subsubsection{Configuring the Finite Energy GKP State}

From the preliminary analysis, the candidate optical GKP state in Eq.(\ref{20}) is found to be closer to the negative form of $\psi_{\text{GKP}}^{-}(x)$. As shown in Fig.\ref{fig:5}, Eq.(\ref{20}) produces only three dominant peaks. Therefore, before evaluating the fidelity parameter, the finite-energy GKP state must be configured to exhibit only three dominant peaks. This configuration is achieved by adjusting the parameter $\Delta$ in Eq.(\ref{5}) and Eq.(\ref{6}). Increasing $\Delta$ reduces the number of peaks in the finite-energy GKP state, since $\Delta$ determines the width of the Gaussian envelope. The presence of this Gaussian envelope introduces a perturbation effect on the finite-energy GKP state, which can be understood through perturbation theory in quantum mechanics. The physical consequences of this perturbation effect are discussed in detail in Appendix~B. Based on these considerations, the optimal choice for achieving approximately three dominant peaks in Eq.(\ref{5}) and Eq.(\ref{6}) is $\Delta = 0.45$. 
This value is preferred because it preserves consistent interpeak spacing in the GKP states. Configuring the finite-energy GKP state to only three dominant peaks, however, represents a limitation of this research.

\begin{equation}
    \label{24}
    \begin{split}
        \Delta l_{0} &= |2\sqrt{\pi}-3.64 | \approx0.10 \\
        \Delta l_{1} &=|\sqrt{\pi}-1.82| \approx 0.05 
    \end{split}
\end{equation}

\begin{figure}[ht]
    \centering
    \includegraphics[width=1\linewidth]{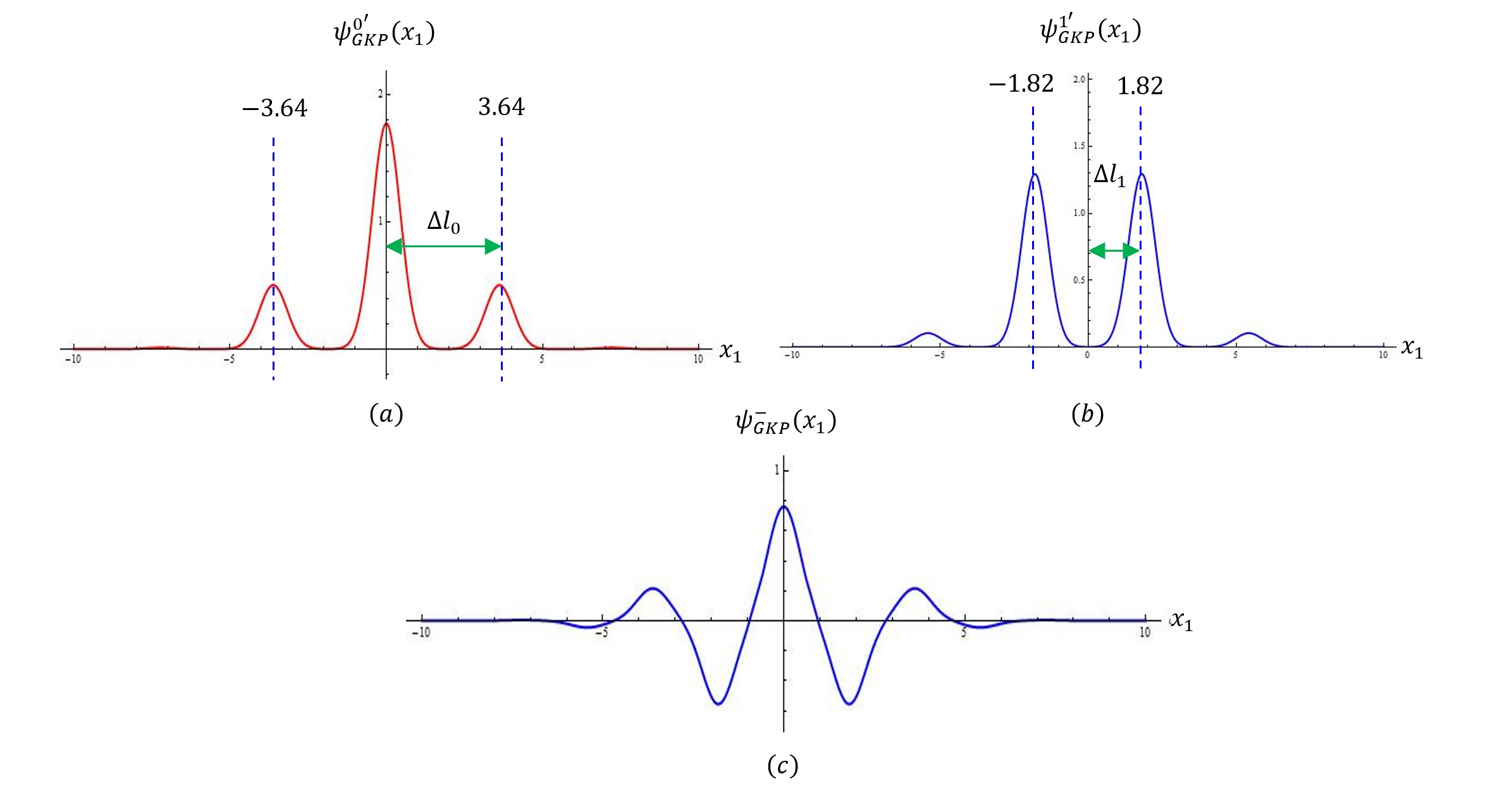}
    \caption{(a) The non-normalized finite Energy GKP State $\psi_{\text{GKP}}^{'0}(x)$ and $\psi_{\text{GKP}}^{'1}(x)$ for $\Delta=0.45$. Meanwhile, (c) is the non-normalized $\psi_{\text{GKP}}^{-}(x)$}
    \label{fig:8}
\end{figure}
For constructing Fig.\ref{fig:8}(a) and Fig.\ref{fig:8}(b), the series in Eq.(\ref{5}) and Eq.(\ref{6}) is truncated at $j = [-10,10]$, resulting in 21 terms. This truncation is sufficient to produce a smooth curve for the finite-energy GKP state, since contributions from higher values of $j$ have negligible impact on the result. From the states $\psi_{\text{GKP}}^{\,'0}(x)$ and $\psi_{\text{GKP}}^{\,'1}(x)$, shown in Fig.\ref{fig:8}(a) and Fig.\ref{fig:8}(b), one can construct $\psi_{\text{GKP}}^{-}(x)$, as illustrated in Fig.\ref{fig:8}(c). 
Eq.(\ref{20}) is then compared with this $\psi_{\text{GKP}}^{-}(x)$ to determine the optimal squeezing parameter $r$ 
through fidelity calculation.

\subsubsection{Fidelity of the Optical GKP State from Single-Photon-Added Squeezed Vacuum}
In the fidelity calculation for Eq.(\ref{20}), the comparison is made with $\psi_{\text{GKP}}^{-}(x)$ using $\Delta = 0.45$ and the definition in Eq.(\ref{23}). 
For this analysis, the indices $n$ and $n'$ are restricted to the range $[0,8]$. This truncation is chosen because values beyond $n = n' = 8$ significantly increase computational time while contributing negligibly to the result. To ensure precision in the fidelity evaluation of Eq.(\ref{20}), the normalization constant $N_{\text{GKP}}^{-}$ is introduced for $\psi_{\text{GKP}}^{-}(x)$. The fidelity calculation then proceeds by first evaluating the integrand of Eq.(\ref{23}).
 
\begin{equation}
    \label{25}
    \tilde{\psi}^{*}_{\text{GKP}}(x_{1})(-\psi_{\text{GKP}}^{-}(x_{1}))=-\frac{N^{-}_{\text{GKP}}N_{\text{S}}\sqrt{2}}{\pi I_{0}\Delta}e^{-\frac{1}{2}x_{1}^{2}}\left[B_{1}(x_{1})-B_{2}(x_{2})\right]
\end{equation}
Here, the functions $B_{1}(x)$ and $B_{2}(x)$ are explicitly dependent on the coordinate $x_{1}$ as defined in Eq.(\ref{26}) and Eq.(\ref{27}). These functions play a role in structuring the mathematical form of the output state and are introduced to capture the dependence of the system on the quadrature variable $x_{1}$.

\begin{equation}
    \label{26}
    B_{1}(x_{1})=\sum^{\infty}_{j=-\infty}C_{j}^{1}(x_{1})\left(\sum^{\infty}_{n,n'=0}\frac{(-1)^{n+n'}\tanh^{n+n'}(-r)}{2^{2n+2n'+1}n!n'!}\varphi_{n,n'}(x_{1})\right)
\end{equation}
\begin{equation}
    \label{27}
    B_{2}(x_{1})=\sum^{\infty}_{j=-\infty}C_{j}^{2}(x_{1})\left(\sum^{\infty}_{n,n'=0}\frac{(-1)^{n+n'}\tanh^{n+n'}(-r)}{2^{2n+2n'+1}n!n'!}\varphi_{n,n'}(x_{1})\right)
\end{equation}
The functions $C_{j}^{1}(x_{1})$ and $C_{j}^{2}(x_{1})$ are defined explicitly in Eq.(\ref{28}) and Eq.(\ref{29}). Both depend on the quadrature variable $x_{1}$, and together they contribute to the mathematical structure required for evaluating the fidelity and characterizing the candidate optical GKP state. 

\begin{equation}
    \label{28}
    C_{j}^{1}(x_{1})=e^{-2\Delta^{2}j^{2}\pi}e^{-\frac{(x_{1}-2j\sqrt{\pi})^{2}}{2\Delta^{2}}}e^{-\frac{1}{2}\Delta^{2}\left[\frac{1}{4}(x_{1}-2j\sqrt{\pi})^{2}-2(x_{1}-2j\sqrt{\pi})j\sqrt{\pi}\right]}
\end{equation}
\begin{equation}
    \label{29}
    C_{j}^{2}(x_{1})=e^{-\frac{\Delta^{2}(2j+1)^{2}\sqrt{\pi}}{2}}e^{-\frac{(x_{1}-(2j+1)\sqrt{\pi})^{2}}{2\Delta^{2}}}e^{-\frac{1}{2}\Delta^{2}\left[\frac{1}{4}(x_{1}-(2j+1)\sqrt{\pi})^{2}-(2j+1)(x_{1}-(2j+1)\sqrt{\pi})\sqrt{\pi}\right]}
\end{equation}
After evaluating the integrand of Eq.(\ref{23}), the next step is to compute the full integral in Eq.(\ref{23}) to obtain the fidelity $F_{1}$. In this context, the integral functions as the overlap integral between two states:  $\tilde{\psi}^{*}_{\text{GKP}}(x_{1})$ and $-\psi_{\text{GKP}}^{-}(x_{1})$. The fidelity expression in Eq.(\ref{30}) contains only one free parameter—the squeezing parameter $r$. To establish the relationship between $r$ and $F_{1}$, the integral in Eq.(\ref{30}) must be calculated, and the resulting values plotted to produce the graph of $F_{1}$ as a function of $r$.

\begin{equation}
    \label{30}
    F=F(r)=\frac{2(N_{\text{GKP}}^{-})^{2}N_{\text{S}}^{2}}{\pi^{2}I_{0}^{2}\Delta^{2}}\left|\int^{\infty}_{-\infty}e^{-\frac{1}{2}x_{1}^{2}}(B_{1}(x_{1})-B_{2}(x_{1}))dx_{1}\right|^{2}
\end{equation}

\begin{figure}[ht]
    \centering
    \includegraphics[width=0.8\linewidth]{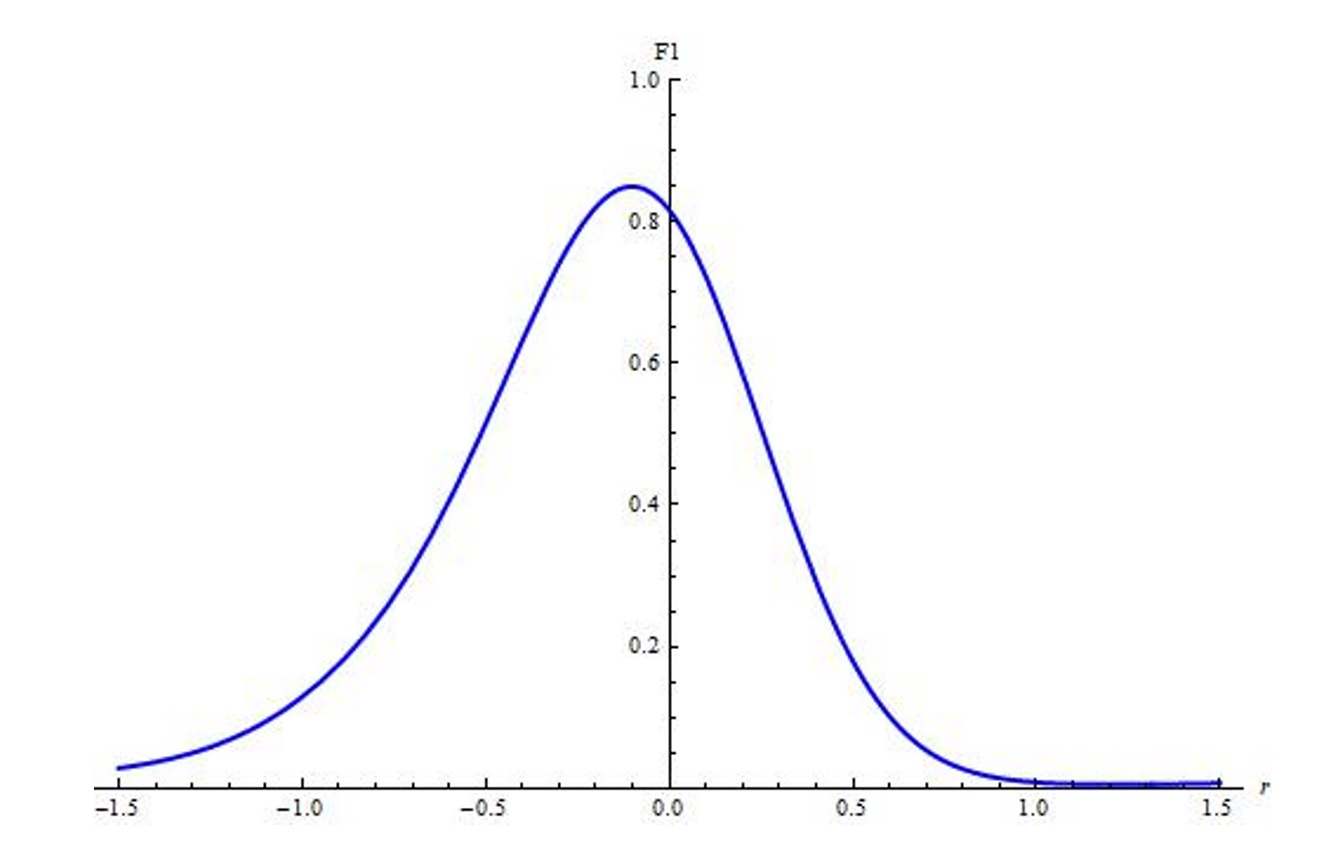}
    \caption{The relationship between the Fidelity $F_{1}(r)=F_{1}$and the parameter $r$}
    \label{fig:9}
\end{figure}

In this research, the integration in Eq.(\ref{30}) is evaluated using the truncation ranges $j=[-10,10]$ and $n=n'=[0,8]$. For the state $-\psi_{\text{GKP}}^{-}(x_{1})$ with $j=[-10,10]$, the calculated ratio between the normalization constant $N_{\text{GKP}}^{-}$ and $\Delta=0.45$ is found to be 0.95. Meanwhile, for the normalization constant $N_{\text{S}}$, a direct calculation method is preferred, since $N_{\text{S}}$ depends explicitly on the squeezing parameter $r$ and directly influences the relationship between the fidelity $F_{1}$ and $r$. Based on the results of these calculations, the dependence of the fidelity $F_{1}$ on the squeezing parameter $r$ is illustrated in Fig.\ref{fig:9}.

\begin{figure}[ht]
    \centering
    \includegraphics[width=1\linewidth]{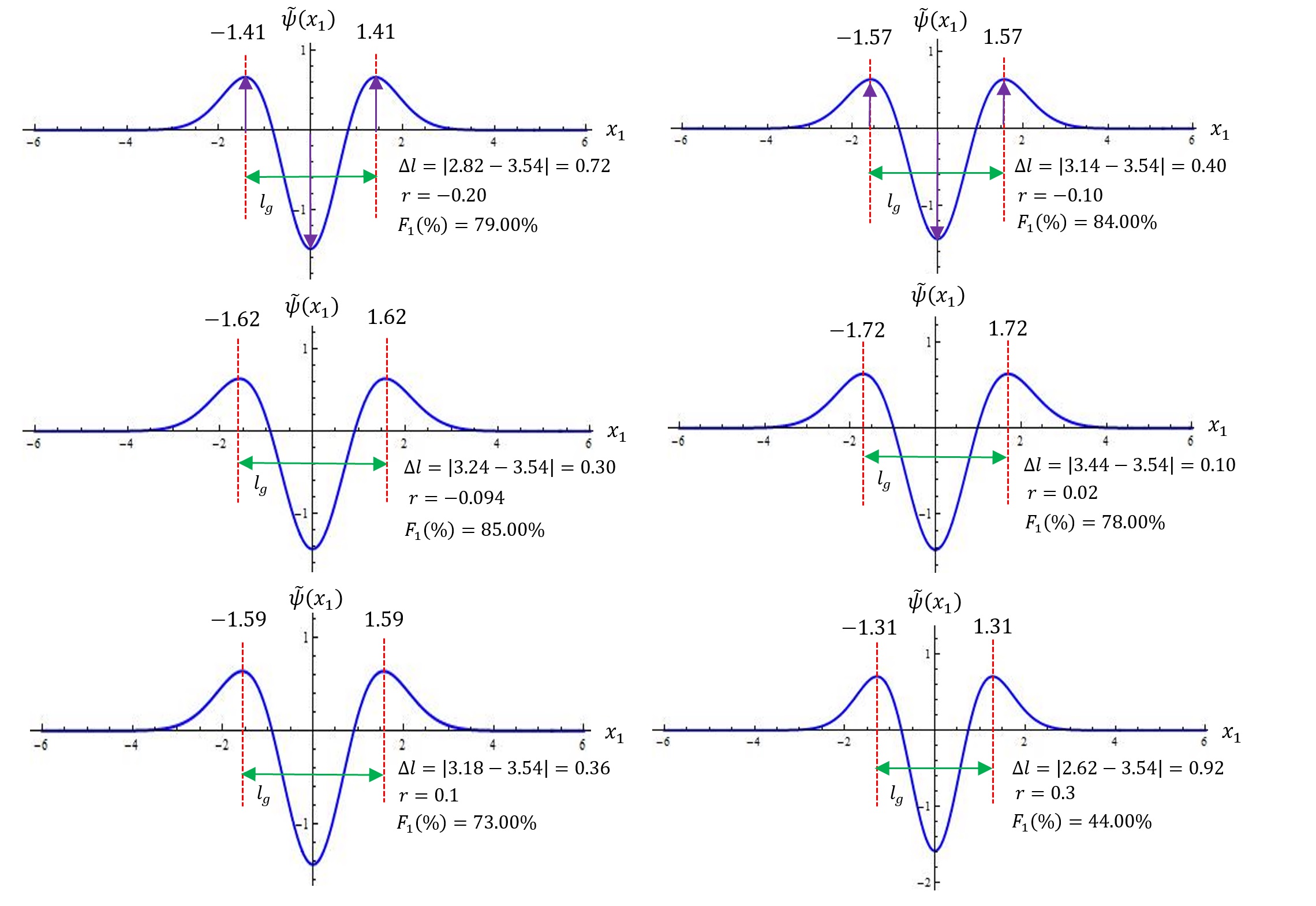}
    \caption{The $\tilde{\psi}(x_{1})$ wave function for different values of $r$. Here, $l_{g}$ is the periodicity of the $\tilde{\psi}(x_{1})$, and $\Delta l$ is the difference between the periodicity of $\tilde{\psi}(x_{1})$ and $-\psi_{\text{GKP}}^{-}(x_{1})$ ($\approx 3.54$)}
    \label{fig:10}
\end{figure}

\begin{figure}[ht]
    \centering
    \includegraphics[width=0.9\linewidth]{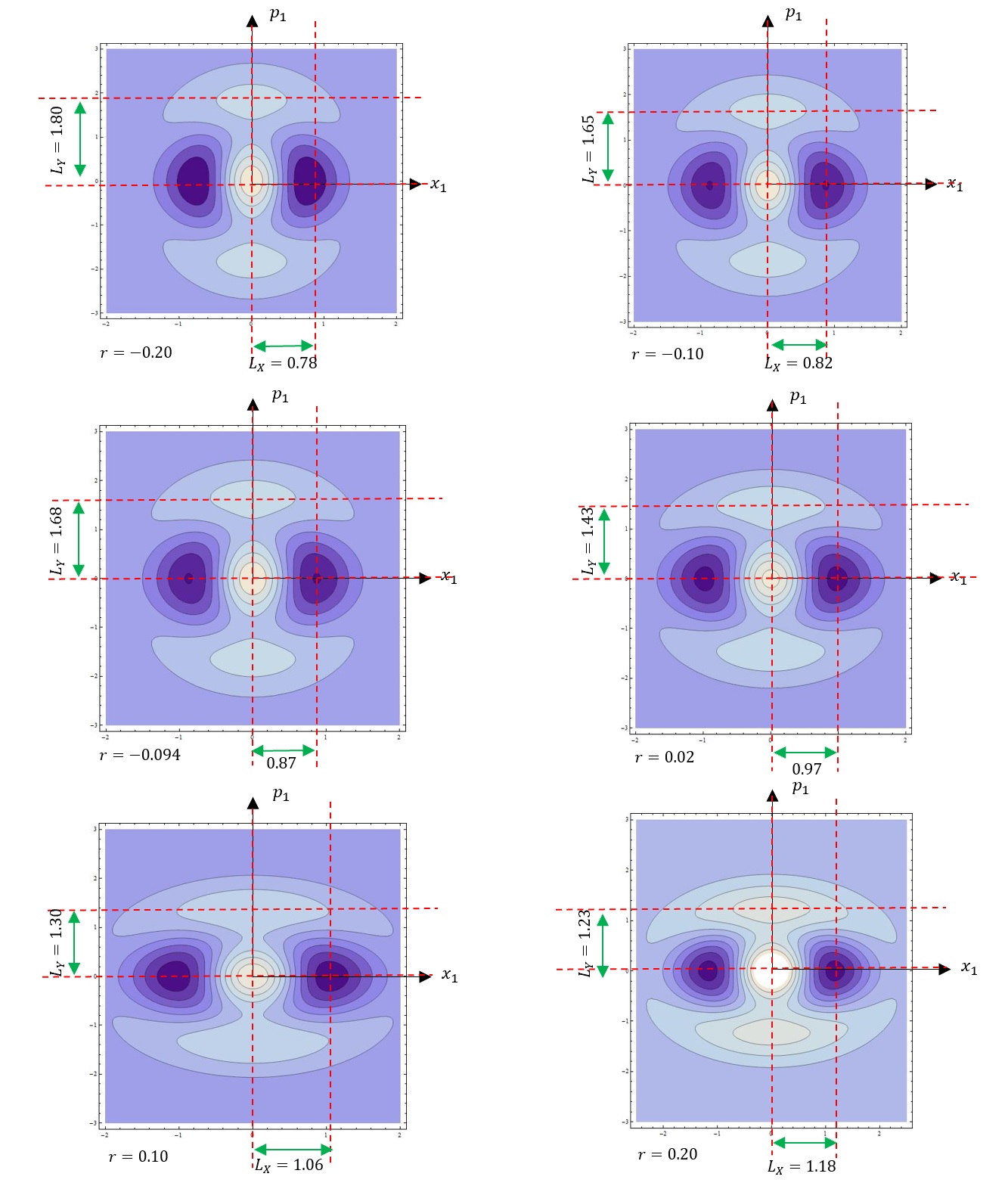}
    \caption{The Wigner function contour of $\tilde{\psi}(x_{1})$ for various $r$ parameters}
    \label{fig:11}
\end{figure}

From Fig.\ref{fig:9}, it is evident that when the squeezing parameter $r$ becomes too large (above 0), the fidelity of Eq.(\ref{20}) decreases. A similar behavior occurs when $r$ falls below -0.10. This reduction in fidelity arises because extreme values of $r$ significantly alter the phase-space quadrature. Such changes in the quadrature modify the periodicity of Eq.(\ref{20}), making it either shorter or longer than the characteristic spacing of $2\sqrt{\pi} \approx 3.54$. As a result, Eq.(\ref{20}) deviates from its resemblance to $-\psi_{\text{GKP}}^{-}(x_{1})$. The impact of this periodicity change is demonstrated by the modifications in both the wave function and the Wigner function of Eq.(\ref{20}) for $r>0$ or $r<-0.10$, as shown in Fig.\ref{fig:10} and Fig.\ref{fig:11}.

According to Fig.\ref{fig:9}, the fidelity $F_{1}$ obtained from Eq.(\ref{20}) remains above $80\%$ $(F_{1} > 0.80)$ 
when the squeezing parameter satisfies $-0.2 < r \leq 0$. 
In terms of the dB scale, this corresponds to $(0 \leq S < 8)\,\text{dB}$. 
The maximum fidelity achieved is $85\%$, occurring at $r = -0.09$. 
This value of $r$ corresponds to a squeezing level of $3.76\,\text{dB}$, 
indicating that Eq.(\ref{20}) most closely approximates $-\psi_{\text{GKP}}^{-}(x_{1})$ under this condition. Importantly, achieving a squeezing level sufficient to obtain fidelity above $80\%$ is experimentally feasible. For example, Nehra and collaborators demonstrated $3.8\,\text{dB}$ vacuum squeezing using few-cycle all-optical measurements in a lithium niobate (LN) nanophotonic platform \cite{Nehra2022}, while Takanashi reported $3.0\,\text{dB}$ squeezing with a semi-monolithic optical parametric oscillator (OPO) \cite{Takanashi2019}, and further results are presented in \cite{Suzuki2006}. These experimental achievements reinforce the conclusion that generating the optical GKP state via a single-photon–added squeezed vacuum is realistically achievable. From the Wigner function contour in Fig.\ref{fig:11}, when the fidelity reaches its maximum at $r = -0.09$, the distance between the central and right circle $(L_{\text{X}})$ and the distance between the central and top circle $(L_{\text{Y}})$ closely resemble those of the Wigner function corresponding to the state $|-\rangle = \frac{1}{\sqrt{2}}\big(|0\rangle_{\text{GKP}} - |1\rangle_{\text{GKP}}\big)$, as reported in \cite{Alvarez2019}. This similarity provides strong evidence that Eq.(\ref{20}), with $r = -0.09$ (equivalent to $3.76\,\text{dB}$), indeed represents the optical GKP state.

\begin{figure}[ht]
    \centering
    \includegraphics[width=1\linewidth]{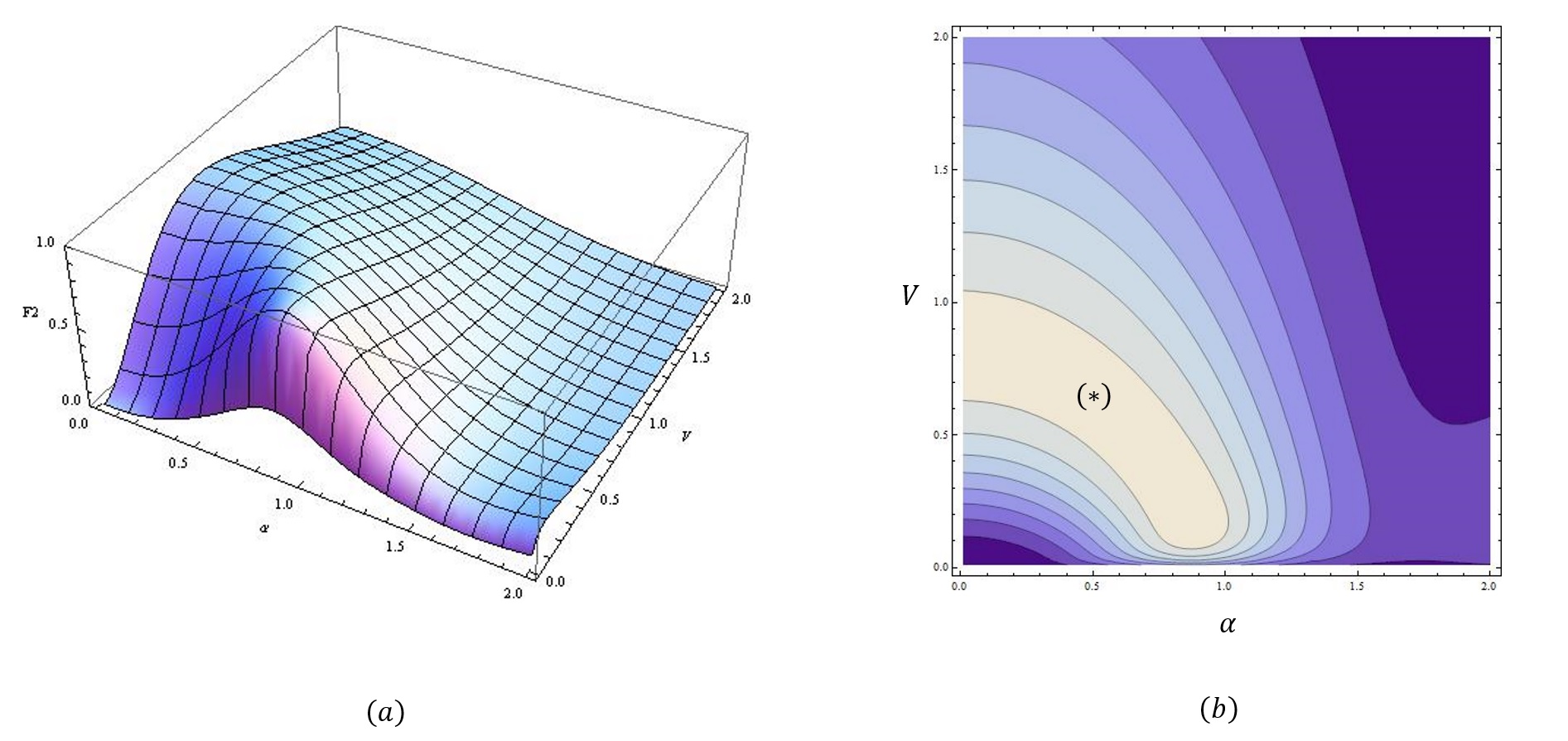}
    \caption{(a) The fidelity surface of the Optical GKP State $\psi_{\text{GKP}}^{-}(x_{1})$ in \cite{Vasconcelos2010} method via the squeezed optical odd-Schrodinger Cat State, and (b) The Fidelity contour plot between the $\alpha$ and $V$ parameters}
    \label{fig:12}
\end{figure}

Applying the single-photon–added squeezed vacuum as the input state to construct the optical GKP state in Eq.(\ref{20}) offers advantages beyond the ease of generation with a modest squeezing level ($3.76\,\text{dB}$). A further benefit is that this approach achieves higher fidelity compared to the method in \cite{Vasconcelos2010}, 
while requiring only a single parameter, $r$. In contrast, the \cite{Vasconcelos2010} method employs the squeezed optical odd Schrödinger cat state, 
which depends on two parameters ($r$ and $\alpha$), and reaches only a maximum fidelity of $80\%$. This difference is evident from the fidelity calculation of the \cite{Vasconcelos2010} method using Eq.(\ref{31}), where the fidelity is plotted as a function of $\alpha$ and $V$ in Fig.\ref{fig:12}.

\begin{equation}
    \label{31}
    F_{2}(\alpha, V)=\frac{(N_{\text{GKP}}^{-})^{2}}{\pi^{3/2}\Delta^{2}V^{1/2}}\left[\frac{1}{3+e^{-4\alpha^{2}/V}-4e^{-\alpha^{2}/V}}\right]\left|\int^{\infty}_{-\infty}(g_{1}(x_{1})-g_{2}(x_{1}))dx_{1}\right|^{2}
\end{equation}
In this formulation, $V=e^{-2r}$ serves as the squeezing parameter, directly linking the fidelity analysis to the degree of squeezing applied. The functions $g_{1}(x_{1})$ and $g_{2}(x_{1})$, defined in Eq.(\ref{32}), depends explicitly on the quadrature variable $x_{1}$.

\begin{equation}
    \label{32}
    \begin{split}
        g_{1}(x_{1}) & = \sum^{\infty}_{j=-\infty} C^{1}_{j}(x_{1})\int^{\infty}_{-\infty}(G(x_{1},V,2\alpha)-2G(x_{1},V,0)+G(x_{1},V,-2\alpha))dx_{1} \\
        g_{2})(x_{1}) &= \sum^{\infty}_{j=-\infty} C^{2}_{j}(x_{1})\int^{\infty}_{-\infty}(G(x_{1},V,2\alpha)-2G(x_{1},V,0)+G(x_{1},V,-2\alpha))dx_{1}
    \end{split}
\end{equation}
In Eq.(\ref{32}), the function $G(x_{1},V,U)=e^{-\frac{1}{2V}(x_{1}-U)^{2}}$ represents a Gaussian function.

Based on Fig.\ref{fig:12}(b), the maximum fidelity of $80\%$ is achieved within the $(*)$ zone. The boundaries of this zone are defined by $0 < \alpha \leq 1.02$ and $0.07 \leq V \leq 1.04$. The range of $V$ corresponds to squeezing levels in the dB scale of $(-41.68 \leq S \leq -2.88)\,\text{dB}$. This result indicates that constructing the optical GKP state $\psi_{\text{GKP}}^{-}(x_{1})$ using the squeezed optical odd Schrödinger cat state requires extremely high levels of squeezing compared to Eq.(\ref{20}), $\tilde{\psi}(x_{1})$. Furthermore, the $\alpha$ values in the \cite{Vasconcelos2010} method fall within the regime of the kitten state, which is experimentally challenging to realize. Difficulties such as mode purity and dark-count probabilities in photon-number detectors hinder the achievement of Wigner function negativity for kitten states \cite{Song2013}. Consequently, the kitten state is insufficient to offset the requirement of immense squeezing in the \cite{Vasconcelos2010} method. 
This reinforces the argument that generating the optical GKP state via a single-photon–added squeezed vacuum 
is more efficient than using the squeezed optical odd Schrödinger cat state. Additionally, employing the single-photon–added squeezed vacuum provides a viable alternative to the Furusawa method, 
which relies on single-photon subtraction \cite{Takase2021}.

\section{Conclusion}
The theoretical analysis presented in this research demonstrates the feasibility of generating the optical GKP state via a single-photon–added squeezed vacuum. 
The specific state achievable is $\psi_{\text{GKP}}^{-}$, with a maximum fidelity of $85\%$ at $r = -0.09$, corresponding to a squeezing level of $3.76\,\text{dB}$. 
This fidelity surpasses the maximum of $80\%$ obtained through the method of \cite{Vasconcelos2010}, which relies on the squeezed optical odd Schrödinger cat state to generate the same $\psi_{\text{GKP}}^{-}$. 
The efficiency of the single-photon–added squeezed vacuum arises from its reliance on a single parameter, $r$, 
whereas the approach in \cite{Vasconcelos2010} requires both $\alpha$ and $r$, with the latter unable to compensate for the need for immense squeezing. 
Consequently, the generation of the optical GKP state $\psi_{\text{GKP}}^{-}$ via a single-photon–added squeezed vacuum represents a more practical and efficient pathway. 

Beyond the immediate results, this work opens several future prospects. First, the modest squeezing requirement (3--4 dB) aligns with current experimental capabilities, suggesting that near-term demonstrations of optical GKP states are realistic. Second, the single-parameter dependence simplifies both theoretical modeling and experimental implementation, providing a scalable route toward larger GKP codewords. Third, integrating this approach with photonic cluster-state architectures may accelerate the development of fault-tolerant quantum computing. Finally, further exploration of multi-photon–added squeezed vacua and hybrid encoding strategies could extend fidelity beyond the current $85\%$, strengthening the robustness of optical GKP states against noise and loss. Taken together, these findings reinforce the role of single-photon–added squeezed vacuum as not only a feasible method for generating optical GKP states, but also as a promising foundation for advancing photonic-based quantum error correction and scalable quantum computing architectures.

\ack{The author gratefully acknowledges Acknowledgement is extended to Alfian Gunawan and Darian Jody Handitya from the Department of Physics, National Tsing Hua University, for careful proofreading and stimulating discussions that contributed to the improvement of this work. Gratitude is also expressed to Muhammad Ismail Yunus from the Department of Mathematics, Institut Teknologi Bandung, for stimulating discussions on the mathematical formulation.
}

\section*{Appendix}
The appendices provide detailed derivations that support the main analysis. These derivations cover three key subjects. First, Appendix A presents the coordinate transformation arising from the 50:50 beam splitter operation. Second, Appendix B applies perturbation theory in quantum mechanics to explain how the width of the Gaussian envelope influences the finite-energy GKP state. Together, these derivations provide the mathematical foundation for the results discussed in the main text.

\subsection*{Appendix A: Coordinate Transformation in the 50:50 Beam Splitter Operation}
\label{APA}
The 50:50 beam splitter operator  $\hat{U}_{\text{BS}}(\pi/4)$ is expressed in Eq.(\ref{18}) as a $2\times 2$ matrix. This matrix is unitary. To demonstrate the coordinate transformation $\hat{x}_{1}\rightarrow \frac{1}{\sqrt{2}}(\hat{x}_{1}+\hat{x}_{2})$ and $\hat{x}_{2}\rightarrow\frac{1}{\sqrt{2}}(\hat{x}_{1}-\hat{x}_{2})$ produced by the 50:50 beam splitter, see Fig.\ref{fig:13} below.

\begin{figure}[ht]
    \centering
    \includegraphics[width=0.4\linewidth]{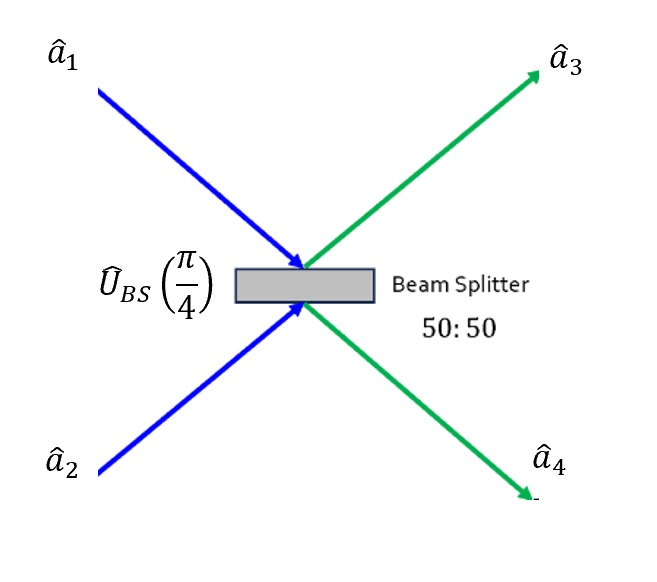}
    \caption{The 50:50 Beam Splitter}
    \label{fig:13}
\end{figure}
\noindent By applying the beam splitter operator $\hat{U}_{\text{BS}}(\pi/4)$ from Eq.(\ref{18}), the annihilation operators at the input ports $(\hat{a}_{1},\hat{a}_{2})$ are transformed into the annihilation operators at the output ports $(\hat{a}_{3},\hat{a}_{4})$.

\begin{equation}
\label{A1}
\begin{bmatrix}
\hat{a}_3 \\
\hat{a}_4
\end{bmatrix}
= \hat{U}_{\text{BS}}(\pi/4)
\begin{bmatrix}
\hat{a}_1 \\
\hat{a}_2
\end{bmatrix}
= \frac{1}{\sqrt{2}}
\begin{bmatrix}
1 & 1 \\
1 & -1
\end{bmatrix}
\begin{bmatrix}
\hat{a}_1 \\
\hat{a}_2
\end{bmatrix}
\tag{A.1}
\end{equation}

Based on Eq.(\ref{A1}), both \(\hat{a}_3\) and \(\hat{a}_4\) are linear combinations of \(\hat{a}_1\) and \(\hat{a}_2\).  
Using the definition of the annihilation and creation operators, one can write
$\hat{x} = \frac{1}{\sqrt{2}} \left( \hat{a} + \hat{a}^\dagger \right)$. Therefore, Eqs.(\ref{A2}) and (\ref{A3}) follow as the coordinate transformations to be proven:

\begin{equation}
\label{A2}
\hat{x}_3 = \frac{1}{\sqrt{2}} \left( \hat{a}_3 + \hat{a}_3^\dagger \right)
= \frac{1}{\sqrt{2}} \left( \tfrac{1}{\sqrt{2}}\hat{a}_1 + \tfrac{1}{\sqrt{2}}\hat{a}_2 
+ \tfrac{1}{\sqrt{2}}\hat{a}_1^\dagger + \tfrac{1}{\sqrt{2}}\hat{a}_2^\dagger \right)
= \frac{1}{\sqrt{2}} \left( \hat{x}_1 + \hat{x}_2 \right)
\tag{A.2}
\end{equation}
\begin{equation}
\label{A3}
    \hat{x}_4 = \frac{1}{\sqrt{2}} \left( \hat{a}_4 + \hat{a}_4^\dagger \right)
= \frac{1}{\sqrt{2}} \left( \tfrac{1}{\sqrt{2}}\hat{a}_1 - \tfrac{1}{\sqrt{2}}\hat{a}_2 
+ \tfrac{1}{\sqrt{2}}\hat{a}_1^\dagger - \tfrac{1}{\sqrt{2}}\hat{a}_2^\dagger \right)
= \frac{1}{\sqrt{2}} \left( \hat{x}_1 - \hat{x}_2 \right)
\tag{A.3}
\end{equation}

\subsection*{Appendix B: Perturbation Theory in Quantum Mechanics Point of View of Finite Energy GKP State}

Sec.\ref{resdis} explains that increasing the width of the Gaussian envelope decreases the number of peaks in the finite-energy GKP state.  
This phenomenon can be described using perturbation theory in quantum mechanics.  
The ideal GKP state Hamiltonian in Eq.(\ref{2}) exhibits two-fold degeneracy with energy \(-2J_{o}\).  
Introducing the Gaussian envelope to normalize \(\psi_{\text{GKP}}^{0}(x)\) and \(\psi_{\text{GKP}}^{1}(x)\) automatically induces a perturbation to the initial system.  
To show this perturbation, we begin with Eq.(\ref{A4}) and Eq.(\ref{A5}):

\begin{equation}
    \label{A4}
    \hat{S}_i' = e^{-\Delta^2 \hat{n}} \, \hat{S}_i \, e^{\Delta^2 \hat{n}}, 
\quad i \in \{X,Z\}
\tag{A.4}
\end{equation}
\begin{equation}
    \label{A5}
    \psi_{\text{GKP}}^{(i')} (x) = N_i \, e^{-\Delta^2 \hat{n}} \, \psi_{\text{GKP}}^{i}(x), 
\quad i \in \{X,Z\}
\tag{A.5}
\end{equation}

Eq.~(A.4) and Eq.~(A.5) approximate the stabilizer operators and the GKP state wave function under the presence of a Gaussian envelope of width \(\Delta\), with \(\hat{n} = \hat{a}^\dagger \hat{a}\) as the number operator \cite{Grismo2021}.  
These approximations preserve the properties of the ideal GKP state, as shown in Eq.(\ref{A6}):

\begin{equation}
    \label{A6}
    \tag{A.6}
\begin{split}
\hat{S}_i' \, \psi_{\text{GKP}}^{(i')} (x) 
&= N_i \left( e^{-\Delta^2 \hat{n}} \hat{S}_i e^{\Delta^2 \hat{n}} \right) 
\left( e^{-\Delta^2 \hat{n}} \psi_{\text{GKP}}^{i}(x) \right) = (+1)\,\psi_{\text{GKP}}^{(i')} (x), 
\quad i \in \{X,Z\}\
\end{split}
\end{equation}

\noindent From Eq.(\ref{A4}), the GKP state Hamiltonian can be rewritten as:
\begin{equation}
    \label{A7}
    \hat{H}' = -\tfrac{1}{2} J_{o} 
\left[ \hat{S}_X' + \hat{S}_Z' + \hat{S}_X^{(+')} + \hat{S}_Z^{(+')} \right]
\tag{A.7}
\end{equation}

\noindent If the Gaussian envelope width is sufficiently small (\(\Delta < 1\)), the exponential term can be expanded into the first two terms of its Taylor series. Eq.(\ref{A7}) then becomes:

\begin{equation}
    \label{A8}
    \hat{H}' = \hat{H} - \tfrac{1}{2} J_{o} \Delta^2 \hat{\alpha}
\tag{A.8}
\end{equation}
where $
\hat{\alpha} = [\hat{S}_X, \hat{n}] + [\hat{S}_Z, \hat{n}] + [\hat{n}, \hat{S}_Z^+] + [\hat{n}, \hat{S}_X^+]$.

The last term in Eq.(\ref{A8}) shows mathematically that the Gaussian envelope acts as a perturbation to the initial system.  
According to perturbation theory for degenerate cases, one constructs the perturbation matrix \(\hat{H}_1\)~\cite{Griffiths2018, Gasiorowicz2003, Sakurai2017}, with elements $
H_{jk} = \langle \psi_j | \hat{H}_1 | \psi_k \rangle$. To obtain the first-order energy correction, Eq.(\ref{A9}) is used:

\begin{equation}
    \label{A9}
    \det \left| \hat{H}_1 - E^{(1)} \hat{I} \right| 
= \det \left|
\begin{array}{cccc}
H_{11} & H_{12} & \cdots & H_{1k} \\
H_{21} & H_{22} & \cdots & H_{2k} \\
\vdots & \vdots & \ddots & \vdots \\
H_{j1} & H_{j2} & \cdots & H_{jk}
\end{array}
- E^{(1)} \hat{I} \right| = 0
\tag{A.9}
\end{equation}
Since the ideal GKP state is a two-fold degenerate system (\(\psi_1 = \psi_{\text{GKP}}^{0}(x)\), \(\psi_2 = \psi_{\text{GKP}}^{1}(x)\)), the perturbation matrix \(\hat{H}_1\) has dimension \(2 \times 2\). Thus, Eq.(\ref{A9}) reduces to:

\begin{equation}
    \label{A10}
    \det \left|
\begin{array}{cc}
H_{11} - E^{(1)} & H_{12} \\
H_{21} & H_{22} - E^{(1)}
\end{array}
\right| = 0
\tag{A.10}
\end{equation}
The first-order energy corrections are then given by:

\begin{equation}
    \label{A11}
    E_{1,2}^{(1)} = \tfrac{1}{2} \left( H_{11} + H_{22} \right) 
\pm \tfrac{1}{2} \sqrt{ \left( H_{11} + H_{22} \right)^2 - \left( H_{11}H_{22} - H_{12}H_{21} \right) }
\tag{A.11}
\end{equation}

\begin{figure}[ht]
    \centering
    \includegraphics[width=1\linewidth]{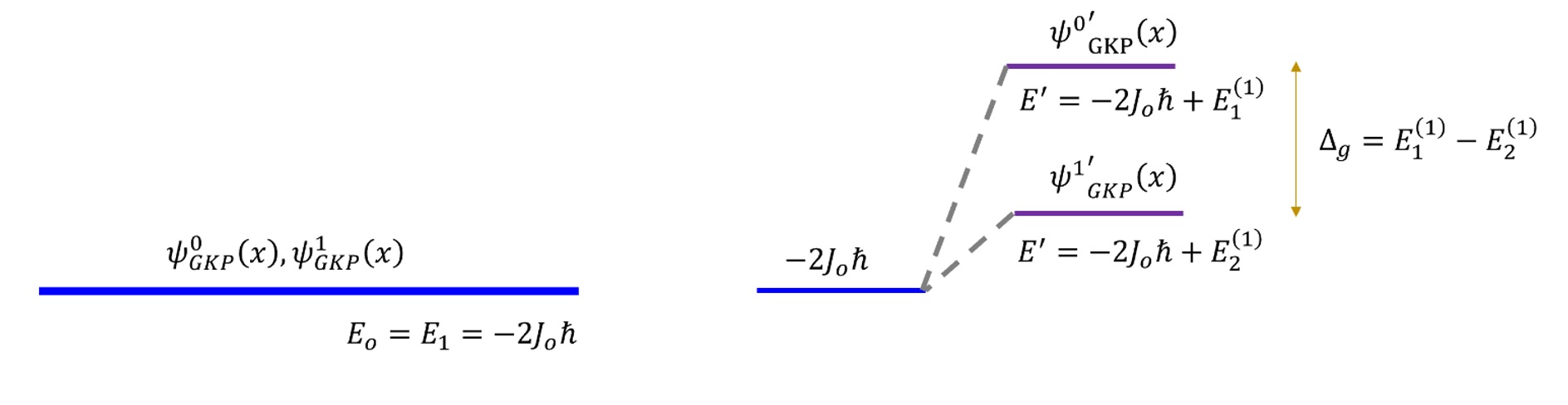}
    \caption{The pictorial representation of the energy gap in the GKP state}
    \label{fig:14}
\end{figure}

\bibliography{bibliography}
\bibliographystyle{iopart-num}

\end{document}